\documentclass[aps,prb,reprint,superscriptaddress,amsmath,amssymb]{revtex4-2}
\usepackage{graphicx} 
\usepackage{float}
\usepackage{amsfonts,amsmath,bbm}
\usepackage[colorlinks=true,linkcolor=blue]{hyperref}
\usepackage{physics}
\usepackage{color}
\usepackage{soul}
\usepackage{mathtools}
\usepackage{booktabs}
\usepackage[normalem]{ulem}
\usepackage{cancel}
\usepackage{minted}

\begin{document}

\title{Extracting the full conductivity tensor in a rectangular sample}
\author{Julia Gelfond}
\email{jgelfon1@jh.edu}
\affiliation{William H. Miller III Department of Physics and Astronomy, Johns Hopkins University, Baltimore, Maryland, 21210}
\affiliation{William I. Fine Theoretical Physics Institute, University of Minnesota, Minneapolis, Minnesota, 55455}
\author{Oskar Vafek}
\email{vafek@umn.edu}
\affiliation{William I. Fine Theoretical Physics Institute, University of Minnesota, Minneapolis, Minnesota, 55455}

\date{\today}

\begin{abstract}
   Electrical transport measurements reveal that many 2D materials exhibit anisotropic conductivity. However, current methodologies for rectangular geometries can only extract a partial conductivity tensor or are difficult to execute experimentally. Here, we propose a simple experimental procedure to extract the full conductivity tensor (or the resistivity tensor by inversion) including the principal axes angle. Our procedure is developed by using a conformal mapping approach to obtain an analytical expression for the potential with a point source and drain on the perimeter. Our solution agrees very well with numerical simulations using COMSOL and the known limiting case where the principal axes angle vanishes. Finite source/drains can be modeled using superposition.
\end{abstract}

\maketitle

\section{Introduction}
Several recent classes of 2D, or layered, materials such as unconventional superconductors \cite{Wu2020Electronic}\cite{Wu2017Spontaneous}, chiral tellurium \cite{Suarez-Rodriguez2024Odd}, rhombohedral graphene \cite{Qin2025Stripe}, and GaAs in the quantum Hall stripe phase \cite{Sammon2019Resistivity} \cite{Fu2020Hidden}\cite{Fogler1996Ground} exhibit anisotropy in magnetoresistance measurements. Extracting the full conductivity tensor, or the resistivity tensor by inversion, from these materials is necessary in order to fully characterize their transport properties. Current methods to do so such as the sunflower geometry \cite{Suarez-Rodriguez2024Odd}\cite{Qin2025Stripe}\cite{Vafek2023Anisotropic}\cite{Zhang2026Angular}\cite{Kang2019Nonlinear} are not always convenient for materials that are more easily fabricated in a rectangular geometry. One can extract the full conductivity tensor for a rectangular sample by using a sunbeam configuration \cite{Wu2020Electronic}\cite{Wu2017Spontaneous}, but the sunbeam configuration is difficult and time-intensive to fabricate. Thus, a simpler procedure is desirable.

Historically, conformal mapping has provided approaches to determine the conductivity tensor in 2D materials. For instance, van der Pauw's 1958 method is commonly used to extract the partial resistivity tensor for arbitrarily shaped flat materials \cite{vanderPauw1958A}. However, it was developed for isotropic materials and cannot separate anisotropic conductivities. Doing so is of great current interest in 2D materials, especially in moiré or low symmetry systems. More recently, using conformal mapping, Peng et al. developed a numerical method to extract the resistivity tensor from an arbitrarily shaped 2D flake using conformal mapping \cite{Peng2018All}. Their technique does not account for Hall effects, whether induced by a perpendicular magnetic field or by spontaneous breaking of time reversal symmetry as in the anomalous Hall effect. Therefore, Peng et al.'s \cite{Peng2018All} method is not sufficient to address current questions when the Hall effect is present for certain materials like GaAs heterostructures. It is an open question why significant deviations from expected resistivity ratios \cite{Lilly1999Evidence} \cite{Simon1999Comment} have been observed
in the quantum Hall stripe phase of GaAs systems at high fractional filling factors \cite{Sammon2019Resistivity}\cite{Fu2020Hidden}\cite{Fogler1996Ground}. Principal axes that are misaligned with the edges of the sample has been proposed as a possible explanation \cite{Simon1999Comment}, but the principal axes angle is currently difficult to extract experimentally.

\begin{figure*}
    \centering
    \includegraphics[width=1\linewidth]{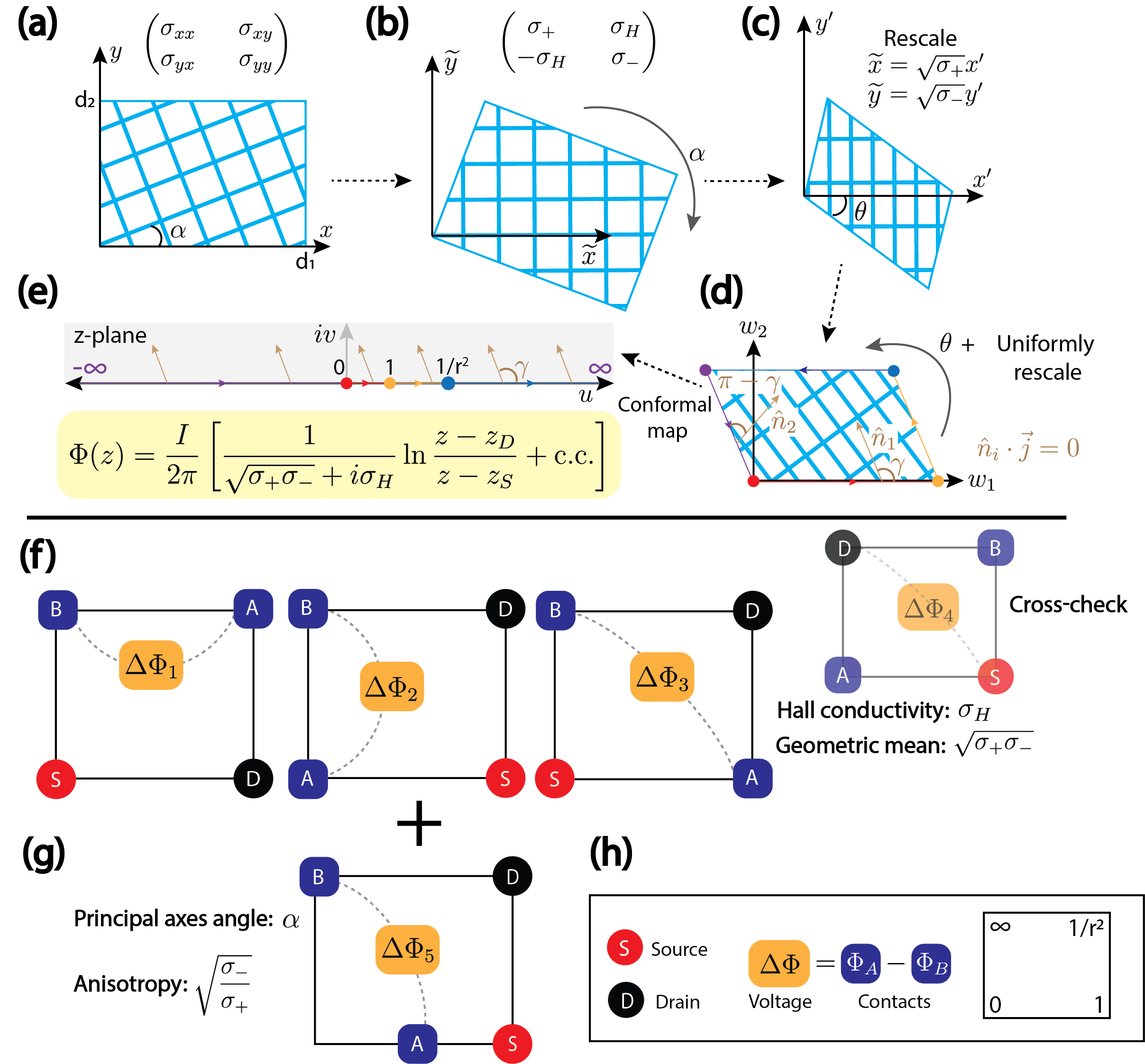}
    \caption{ Analytical solution for electric potential on a rectangle and experimental configurations for extracting the full conductivity tensor and principal axes angle. \textbf{(a)} Original sample with principal axes rotated relative to the lab frame by $\alpha$. \textbf{(b)} Rotation clockwise by $\alpha$. \textbf{(c)} Anisotropic rescaling. \textbf{(d)} Rotation counterclockwise by $\theta$ and  uniform rescaling. \textbf{(e)} Schwarz-Christoffel conformal map and analytical solution for the potential in the complex z-plane. \textbf{(f)} Three vertex configurations which together give the Hall conductivity and geometric mean of the longitudinal conductivities. The fourth vertex configuration does not give new information, but can be used as a cross-check. \textbf{(g)} Midpoint configuration that combined with the corner configurations gives the principal axes angle and anisotropy, therefore the entire conductivity tensor.  \textbf{(h)} Key for experimental configurations.}
    \label{abstract}
\end{figure*}

In this paper, we use a conformal mapping approach to derive an analytical solution to the electric potential in a rectangular sample. We then propose a simple experimental procedure with four measurements to extract the full conductivity tensor (including Hall components) and the principal axes angle using our analytical solution (see Fig.\ref{abstract} for a visual overview). The setup is a rectangular sample of side length $d_1$ along the x-axis and $d_2$ along the y-axis with the bottom left corner at the origin. One source and one drain of equal magnitude are attached to the perimeter of the sample. Boundary conditions require that the current is contained to the sample. The sample is taken to be uniform so that the conductivity tensor, $\sigma$, and principal axes angle, $\alpha$, are constant throughout the sample. The principle axis of the sample can be taken to be at an angle $\alpha\in[0,\pi)$ measured counterclockwise from the bottom edge of the sample without loss of generality.

The paper is structured as follows. First, we analytically solve for the potential across the sample for arbitrary source and drain placement along the perimeter using conformal mapping (see Section \ref{analysis}). We provide our solution here for convenience,
\begin{equation}
    \Phi(z) = \frac{I}{2\pi} \left[ \frac{1}{ \sqrt{\sigma_+\sigma_-}  + i \sigma_H} \ln{\frac{z-z_D}{z-z_S}} + \text{c.c.} \right].
\end{equation}
Importantly, note that $z$ is a coordinate in the upper half of the complex plane, not the original coordinates of the rectangle. Here, $I$ is the current, $\sigma_H$ is the Hall conductivity, $\sigma_\pm$ are the anisotropic conductivities in the principal axes frame, and $z_{S/D}$ are the locations of the source/drain along the perimeter of the sample in the complex plane. 
Next, in Section \ref{exp}, we propose an experimental procedure with four configurations to extract the full conductivity tensor and principal axes angle. For convenience, we provide a \href{https://github.com/jag47/Magnetoconductance}{GitHub page} with Python and Mathematica files that automates the formulas to determine $\sigma$ and $\alpha$ from resistance measurements in Section \ref{exp}. Finally, after deriving our solution, we compare our analytical solution to simulated data from COMSOL and find excellent agreement (see Section \ref{discussion}). In the limiting case of vanishing anisotropy angle, our solution agrees with previous literature \cite{Simon1999Comment}.

\section{Analysis} \label{analysis}
    Our main assumption is that Ohm's law holds throughout the sample, 
    \begin{equation} \label{ohm}
    \mathbf{j}_\mu = \sigma_{\mu\nu} \mathbf{E}_\nu = -\sigma_{\mu\nu} \partial_\nu \Phi \qquad \mu,\nu=1,2,
    \end{equation} where $\mathbf{j_\mu}$ is the current density, $\mathbf{E_\nu}$ is the electric field with electric potential $\Phi$, and $\sigma_{\mu\nu}=\begin{pmatrix}\sigma_{xx} && \sigma_{xy} \\ \sigma_{yx} && \sigma_{yy} \end{pmatrix}$ is the conductivity tensor in the original coordinates of the sample where the off diagonal components are not necessarily antisymmetric. We assume the sample is uniform so that the potential is position dependent, but the conductivity matrix is not. Using equation \eqref{ohm}, the continuity equation for a sample with one source at $\mathbf{r}_S$ and one drain at $\mathbf{r}_D$ of equal magnitude, $I$, becomes
    \begin{equation} \label{cont}
    \div{\mathbf{j}} = -\div( \sigma \grad \Phi) = I(\delta(\mathbf{r}-\mathbf{r}_{S})-\delta(\mathbf{r}-\mathbf{r}_{D})).
\end{equation} 

    The boundary conditions are neither Dirichlet or Neumann. Rather, they are oblique---the directional derivative across each edge vanishes \begin{equation} \label{bound}
        \begin{split}
    &\sigma_{xx} \pdv{\Phi}{x} + \sigma_{xy} \pdv{\Phi}{y}=0 \quad \text{at} \quad x=0,x=d_1
    \\
    &\sigma_{yx} \pdv{\Phi}{x} + \sigma_{yy} \pdv{\Phi}{y}=0 \quad \text{at} \quad y=0,y=d_2,
\end{split}
    \end{equation} so that the current is contained to the sample. 

    In order to the solve the differential equation \eqref{cont}, it is convenient to transform the left hand side to the Laplacian. This involves a rotation followed by an anisotropic rescaling i.e. an affine transformation. To cleanly incorporate the boundary conditions, the sample is then rotated, uniformly rescaled, and conformally mapped to the upper half of the complex plane using the Schwarz-Christoffel map. In the complex plane, the solution is easily acquired using Fourier transformations. The details follow in the next couple sections.

    \subsection{Preparing for Conformal Mapping: Affine Transformation to a Parallelogram}
    First, the coordinates are rotated clockwise by angle $\alpha$ so that the principal axes align with the original x and y axes of the sample (see Fig.\ref{abstract}a and b): $\left(\begin{array}{c}\widetilde{x} \\\widetilde{y}\end{array}\right)=R_{-\alpha}\left(\begin{array}{c}x \\y\end{array}\right)$. Here, $R_{\varphi}=\left(\begin{array}{cc} \cos\varphi & -\sin\varphi\\
    \sin\varphi & \cos\varphi\end{array}\right)$
     is the standard counterclockwise rotation matrix by an angle $\varphi$. Antisymmetric components in the derivatives vanish so that the continuity equation \eqref{cont} becomes 
    \begin{equation}
        -\sigma_+ \pdv[2]{\Phi}{\widetilde{x}} - \sigma_- \pdv[2]{\Phi}{\widetilde{y}} = I(\delta(\widetilde{\mathbf{r}}-\widetilde{\mathbf{r}}_S)-\delta(\widetilde{\mathbf{r}}-\widetilde{\mathbf{r}}_{D})),
    \end{equation}
    where $\sigma_+$ and $\sigma_-$ are the diagonal components of the conductivity tensor in the principal axes basis. They relate to the original conductivity matrix via
    \begin{equation}
    \begin{pmatrix}
        \sigma_+ && \sigma_H 
        \\ -\sigma_H && \sigma_-
    \end{pmatrix}= R_{-\alpha}
    \begin{pmatrix}
        \sigma_{xx} && \sigma_{xy} \\ 
        \sigma_{yx} && \sigma_{yy}
    \end{pmatrix}
    R_{\alpha}.
    \end{equation} Without loss of generality, we define $\sigma_+$, $\sigma_-$, and $\alpha$ so that $\sigma_+\geq\sigma_-$.

    Next, rescale the coordinates according to $\widetilde{x} = \sqrt{\sigma_+}x'$ and $\widetilde{y}=\sqrt{\sigma_-}y'$. The sample now looks like a parallelogram (see Fig.\ref{abstract}c). The continuity equation \eqref{cont} contains the sought after Laplacian operator, 
    \begin{equation} \label{contlaplace}
        \laplacian \Phi= -\frac{I}{\sqrt{\sigma_+\sigma_-}}(\delta(\mathbf{r}'-\mathbf{r}'_{S})-\delta(\mathbf{r}'-\mathbf{r}'_{D})),
    \end{equation}
allowing us to apply conformal mapping techniques.

    After the above clockwise coordinate rotation by $\alpha$ and anisotropic rescaling, the boundary conditions are not particularly simplified. However, they can be simplified using conformal mapping, which allows us to obtain an explicit solution taking into account the boundary conditions. As we discuss in detail in the next section, the conformal mapping can be performed analytically using a Schwarz-Christoffel transformation by Anderson et al. \cite{Anderson2000Generalized}. Anderson et al. map a parallelogram to the upper half of the complex plane. They require that the parallelogram lays on the x-axis and that the corners are specified by hypergeometric functions. Therefore, to match our parallelogram with Anderson et al. \cite{Anderson2000Generalized}, it must be rotated counterclockwise by $\theta$ from the configuration illustrated in Fig.\ref{abstract}c to align the bottom edge with the horizontal axis and then uniformly rescaled as shown in Fig.\ref{abstract}d. This uniform rescaling affects neither the form of the continuity equation \eqref{contlaplace} nor the boundary conditions.

    Thus, in order to align the bottom of the parallelogram with the horizontal axis, we perform a counterclockwise rotation by $\theta$, where \begin{equation}
        \tan \theta = \sqrt{\sigma_+/\sigma_-} \tan \alpha.
    \end{equation}
     The boundary conditions for the bottom and the top (horizontal) edges of the parallelogram become 
    \begin{equation} \label{eq:bcs}
        \sigma_H \pdv{\Phi}{x''} -\sqrt{\sigma_+\sigma_-} \pdv{\Phi}{y''} = 0, 
    \end{equation} 
    where $''$ denotes coordinates in the rotated parallelogram frame (i.e. Fig.\ref{abstract}d before the uniform rescaling):
    $\left(\begin{array}{c}x'' \\y''\end{array}\right)=R_{\theta}\left(\begin{array}{c}x' \\y'\end{array}\right)$.
    As illustrated in Fig.\ref{abstract}d, the directional derivative points at an angle $\gamma$ relative to the bottom horizontal edge, such that $\tan \gamma = -\sqrt{\sigma_+\sigma_-}/\sigma_H$. By symmetry, the analogous angle is the same at the top horizontal edge. Relative to the left and right edges, the angle is $\pi-\gamma$. In order to obtain Eq.\eqref{eq:bcs}, we start with Eq.\eqref{bound} and use the chain rule for the aforementioned series of affine transformations. We obtain
    \begin{equation} \label{dirder}
        \left[R_{\alpha}\right]_{ij} \begin{pmatrix}
        \sigma_+ && \sigma_H\\
    -\sigma_H && \sigma_-
    \end{pmatrix}_{jk}
    S_{kl} \left[R_{-\theta}\right]_{lm}
    \pdv{\Phi}{x'_m}=0,
    \end{equation} 
where $i=1$ corresponds to the boundary conditions on the left and right edges, and $i=2$ to the top and bottom edges. As before, $R_\varphi$ is the standard counterclockwise rotation matrix. The matrix 
    $S =  \begin{pmatrix} 
        1/\sqrt{\sigma_+} && 0 
        \\ 0 && 1/\sqrt{\sigma_-}
        \end{pmatrix}$
    accomplishes anisotropic rescaling. Letting $i=2$ and after much simplification we obtain Eq.\eqref{eq:bcs}. The angle relative to the left and right edges is similarly obtained by setting $i=1$.

    \label{conformal condition}
    Because conformal maps preserve relative angles, the boundary conditions at the edges of the parallelogram --i.e. $\hat{n}\cdot{\bf j}=0$ which can be be expressed as the directional derivative of $\Phi$ at the boundary-- can be readily found along the real axis in the complex $z$-plane. Specifically, because the directional derivative is at the same angle relative to both sides of the parallelogram, the directional derivative is constant along the real axis after the Schwarz-Christoffel (conformal) transformation discussed next.
     
    \subsection{Schwarz-Christoffel Map}
    The Schwarz-Christoffel transformation conformally maps a parallelogram to the upper half of the complex plane. It takes the form \cite{Anderson2000Generalized},
    \begin{align} \label{sc}
        w &= f(z) = \frac{\sin \pi a}{2}  \int_0^z \frac{\dd{t}}{t^a (1-t)^{1-a}(1-r^2t)^a}\nonumber
        \\
        &=\frac{z^{1-a}}{2(1-a)}(\sin \pi a)F_1(1-a;1-a,a;2-a;z,r^2z),
    \end{align}
    where $F_1$ is an Appell hypergeometric function \cite{WeissteinAppell}
    \begin{align} \label{Appell}
F_1(\frak{a};\frak{b},\frak{b}';\frak{c};x,y)&=\sum_{m=0}^\infty\sum_{n=0}^\infty \frac{\frak{a}_{m+n}\frak{b}_m\frak{b}'_n}{m!n!\frak{c}_{m+n}}x^my^n,
    \end{align}
    and $\frak{a}_m=\frak{a}(\frak{a}+1)\ldots(\frak{a}+m-1)$.
    The four edges of the parallelogram in the complex $w$-plane map onto the real axis (see Fig.\ref{abstract}d and e). Starting from the bottom left and moving counterclockwise, the four vertices of the parallelogram map onto the points $0$, $1$, $1/r^2$ and $\infty$, respectively, where the parameter $r\in(0,1)$. The location of the four vertices in the complex $w$ plane can be expressed in closed form using hypergeometric functions \cite{Anderson2000Generalized},
    \begin{align} \label{corners}
        &f(0)=0 \quad \text{bottom left},\nonumber
        \\
        &f(1)=\mathcal{K}_a(r)\equiv \frac{\pi}{2} \prescript{}{2}{F}_1(a,1-a,1;r^2) \quad \text{bottom right}, \nonumber
        \\
        &f\left(\frac{1}{r^2}\right)=\mathcal{K}_a(r) + e^{i(1-a)\pi} \mathcal{K}_a(\sqrt{1-r^2}) \quad \text{top right}, \nonumber
        \\
        &f(\infty)= e^{i(1-a)\pi} \mathcal{K}_a(\sqrt{1-r^2}) \quad \text{top left}.
    \end{align}
    The 
     Gaussian hypergeometric function appearing above can be defined via the series \begin{equation}
        \prescript{}{2}{F}_1(\frak{a},\frak{b},\frak{c};z)=\sum_{n=0}^\infty\frac{\frak{a}_n\frak{b}_n}{\frak{c}_n}\frac{z^n}{n!} =1+\frac{\frak{a}\frak{b}}{1!\frak{c}}z + \dots,
    \end{equation}
    which converges for $\abs{z} < 1$ \cite{WeissteinHypergeometric}.
     There is no known analytical expression for the inverse of $f$, and therefore the locations of any other points after the mapping.

    Continuing from the previous section, the sample has been transformed from a rectangle to a parallelogram with one side along the x-axis. To make the sample fit the correct hypergeometric function corner requirements \eqref{corners}, it is uniformly rescaled in the $w$-plane, 
    \begin{equation} \label{unirescale}
        w_1 = \frac{\mathcal{K}_a(r)}{x_1''} x'', \qquad w_2 = \frac{\mathcal{K}_a(r)}{x_1''} y'', 
    \end{equation} where $x_1'' = d_1(\frac{\cos\theta\cos\alpha}{\sqrt{\sigma_+}}+\frac{\sin\theta\sin\alpha}{\sqrt{\sigma_-}}) $ is the coordinate of the bottom right corner of the parallelogram before rescaling. In general, coordinates from the original sample are related to the $''$ (not rescaled) parallelogram coordinates by an affine transformation  
    \begin{equation} \label{pointtransform}
    \begin{pmatrix} 
        x'' \\ y''
    \end{pmatrix}= R_{\theta}SR_{-\alpha} \begin{pmatrix}
        x \\ y
    \end{pmatrix},
\end{equation}
following the same definitions for the transformation matrices from 
Eq.\eqref{dirder}. The continuity equation retains its Laplacian form in the w-plane coordinates,
\begin{equation}
    (\partial_{w_1}^2 + \partial_{w_2}^2)\Phi = - \frac{I}{\sqrt{\sigma_+\sigma_-}}(\delta(\mathbf{w}-\mathbf{w}_S)-\delta(\mathbf{w}-\mathbf{w}_D)),
\end{equation} where $\mathbf{w} = (w_1,w_2)$.

    In Anderson et al. \cite{Anderson2000Generalized}, the hypergeometric parameter, $a$, is restricted so that $a\in(0,1/2]$. In this work, we let {$a\in(0,1)$}. This can be seen by noting that the ratio of the Cartesian coordinates of the top left corner must be equivalent before and after rescaling, giving
    \begin{equation} \label{tana}
    \tan a\pi =  \frac{\sqrt{\sigma_+ \sigma_-}}{\sigma_+ - \sigma_-} \frac{1}{\sin \alpha \cos \alpha}.
    \end{equation}
    It is readily seen from the above equation that if $0<\alpha<\pi/2$ then $a\in(0,1/2)$ and if $\pi/2<\alpha<\pi$ then $a\in(1/2,1)$. In the special case of $a=1/2$,  when the left hand side diverges, $\alpha$ can be either $0$ or $\pi/2$, i.e. the principal axes are aligned with the sample edges. 
    
    Equating the real part of $f(\infty)$ in Eq.\eqref{corners} to $w_1$ in Eq.\eqref{unirescale} with $x''$ determined from Eq.\eqref{pointtransform} (setting $x=0$ and $y=d_2$)
    gives, 
    \begin{equation} \label{ratio}
    \frac{\mathcal{K}_a(r)}{\mathcal{K}_a(\sqrt{1-r^2})} = \frac{d_1}{d_2}\sqrt{\frac{\sigma_-\cos^2 \alpha + \sigma_+\sin^2 \alpha}{\sigma_+ \cos^2 \alpha + \sigma_- \sin^2 \alpha}}.
    \end{equation}
    We will use the above Eq. together with Eq.\eqref{tana} to determine the orientation of the principal axes and the ratio $\sigma_-/\sigma_+$ from resistance measurements shown in  Fig.\ref{abstract}f-h.
    
    Up until now, the source and drain could be anywhere within the sample. Going forward, we assume that the source and drain are infinitesimally close to the perimeter of the sample so that they very nearly map to the real line under the Schwarz-Christoffel transformation.
    Applying the Schwarz-Christoffel map to the continuity equation  \eqref{contlaplace}, yields
    \begin{equation} \label{diffeq}
         \nabla^2 \Phi = -\frac{I}{\sqrt{\sigma_+\sigma_-}} \delta(v-v_{\epsilon})(\delta(u-u_{S}) -\delta(u-u_{D})).
    \end{equation}
    Here $u$ and $v$ are coordinates in the upper half of the complex plane so that $z=u + iv$.  The source and drain are at coordinates $(u_{S}, v_\epsilon \xrightarrow[]{}0^+)$ and $(u_{D}, v_\epsilon \xrightarrow[]{}0^+)$, respectively.

    The boundary conditions \eqref{dirder} are simpler after the conformal map since they become constant in the complex plane (see the comment from the last paragraph of the previous section \ref{conformal condition}), 
    \begin{equation} \label{complex bounds}
    -\sqrt{\sigma_+\sigma_-}\left.\pdv{\Phi}{v}\right|_{v=0} + \sigma_H \left.\pdv{\Phi}{u}\right|_{v=0} =0 \quad \forall u.
    \end{equation}
    
    Using Fourier transforms, the solution for the potential is found to be  (see Appendix \ref{fullsolution} for more details)
    \begin{equation} \label{solution}
    \Phi(z) = \frac{I}{2\pi} \left[ \frac{1}{ \sqrt{\sigma_+\sigma_-}  + i \sigma_H} \ln{\frac{z-z_D}{z-z_S}} + \text{c.c.} \right].
\end{equation}
    This gives us the analytic solution for the electric potential on a 2D rectangular sample for a general conductivity tensor. Note that the source and drain are assumed to be along the perimeter of the sample and that $z$ is a coordinate in the upper half of the complex plane, not the original coordinates of the rectangle. 

    A general point in the z-plane can be mapped back to the original rectangle. Specifically, given $z$, the corresponding w-plane coordinate, $w$, is given by Eq.\eqref{sc}. The corresponding coordinates in the lab frame are then given by  
    \begin{equation}
        \begin{pmatrix}
            x\\y
        \end{pmatrix} = R_\alpha S^{-1} R_{-\theta}\frac{x_1''}{\mathcal{K}_a(r)}
        \begin{pmatrix}
            \mathrm{Re}({w})\\ \mathrm{Im}({w})
        \end{pmatrix}
    \end{equation}
    which invert the series of affine transformations in \eqref{pointtransform} and \eqref{unirescale}. As defined previously, $x_1'' = d_1(\frac{\cos\theta\cos\alpha}{\sqrt{\sigma_+}}+\frac{\sin\theta\sin\alpha}{\sqrt{\sigma_-}})$. The four vertices present an additional advantage in that their images in the z-plane are simply $0$, $1$, $1/r^2$ and infinity. We make use of this fact in the next section.

\section{Procedure for extracting the conductivity tensor} \label{exp}
    We now discuss how to implement our analytical solution \eqref{solution} to extract to the full conductivity tensor and the principal axes angle with four resistance measurements (see the configurations in Fig.\ref{abstract}f-h). Provided with the paper is a \href{https://github.com/jag47/Magnetoconductance}{GitHub page} with Python and Mathematica code implementing the following formulas for finding the materials parameters. 

    Consider the four vertex configurations in Figure \ref{abstract}f. Using our solution \eqref{solution} for configuration 1, the potential for any point in the upper half of the complex $z$-plane is given by 
    \begin{equation}
        \Phi(z) =  \frac{I}{2\pi} \left[ \frac{1}{ \sqrt{\sigma_+\sigma_-}  + i \sigma_H} \ln{\frac{z-1}{z}} + \text{c.c.} \right],
    \end{equation}
    where we plugged in $z_S=0$ as the source and $z_D=1$ as the drain. Across points $A$ and $B$, which map to $1/r^2$ and $\infty$ in the $z$-plane respectively, the potential difference in configuration 1 is
    \begin{align}
        \Delta \Phi_1 &=  \frac{I}{2\pi} \Biggl[ \frac{1}{ \sqrt{\sigma_+\sigma_-}  + i \sigma_H} \ln{\frac{1/r^2-1}{1/r^2}} + \text{c.c.} 
        \\
        &- \left(  \frac{1}{ \sqrt{\sigma_+\sigma_-}  + i \sigma_H} \ln{\frac{\infty-1}{\infty}} + \text{c.c.}\right) \Biggr].
    \end{align}
    The latter logarithm evaluates to $\ln(1)=0$, effectively grounding the sample at the top left vertex.  Factoring out the remaining logarithm in $\Delta\Phi_1$ and simplifying gives  \begin{equation}
        \Delta \Phi_1 =  \frac{I}{\pi} \frac{\sqrt{\sigma_+\sigma_-}}{\sigma_H^2 + \sigma_+ \sigma_-} \ln(1-r^2). 
    \end{equation}
    We note that the pre-factor contains the geometric mean of the resistivity along the two principal axes \begin{equation}
\rho_*=\frac{\sqrt{\sigma_+\sigma_-}}{\sigma_H^2 + \sigma_+ \sigma_-},
\end{equation} 
as follows from the matrix inversion.
    Inspecting configuration 1, the potential is measured across the side with the drain to the side with the source, leading to a negative potential. We define $R_1$ to be $-\Delta\Phi_1/I$ so that $R_1$, a longitudinal resistance, is conventionally positive, \begin{equation} \label{phi1}
        R_1 = -\frac{\rho_*}{\pi} \ln(1-r^2).
    \end{equation}

\begin{figure*}
        \centering
        \includegraphics[width=1\linewidth]{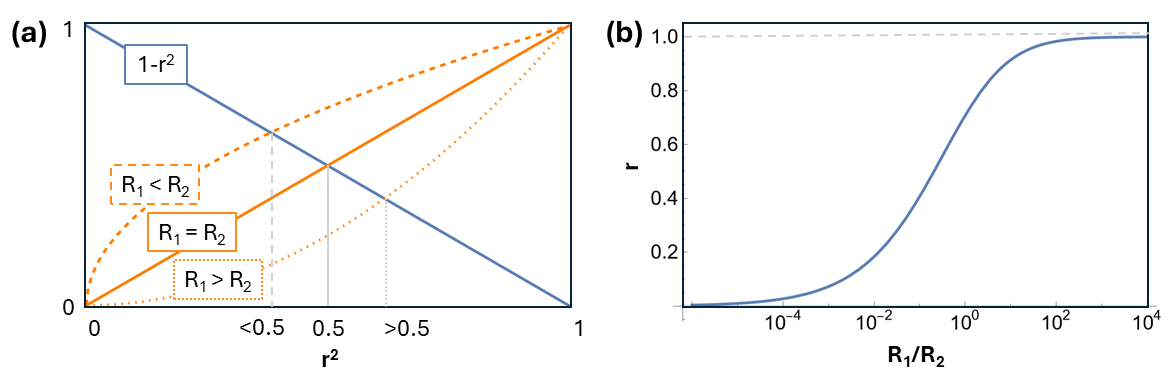}
        \caption{\textbf{(a)} Schematic showing how $1-r^2=r^{2R_1/R_2}$ always has a unique solution for $r^2$. Note that $r^2>0.5$ if $R_1>R_2$ and $r^2<1/2$ if $R_1<R_2$.  \textbf{(b)} Plot of $r$ as a function of $R_1/R_2$ on a base-10 log scale.} 
        \label{r2 eq}
    \end{figure*}

   The potential difference for configuration 2 is found similarly, plugging in $z_S=1$ and $z_D=1/r^2$ for the source and drain and $A=0$ and $B=\infty$ for the contacts,
   $
        \Delta \Phi_2 = I \rho_* \ln(1/r^2)/\pi. 
    $
    Since in configuration 2 point $A$ is closer to the source than point $B$, the potential difference $\Phi_A-\Phi_B$  will be positive. Therefore, there is no need to negate the potential to get the longitudinal resistance, 
     \begin{equation}\label{eq:R2}
        R_2 =  \frac{\rho_*}{\pi} \ln(1/r^2). 
    \end{equation}
    Dividing $R_2$ by $R_1$, and after straightforward algebra, we obtain the equation satisfied by the hypergeometric parameter, $r$,\begin{equation} \label{fixr}
        1-r^2=r^{2R_1/R_2}.
    \end{equation}
Although in general, given the measured value of $R_1/R_2$ this equation needs to be solved numerically for $r$, we can readily understand its solution qualitatively. Consider plotting the left hand side as a function of $r^2$. We obtain a straight line with the slope $-1$, intersecting the horizontal and vertical axes at $1$. The right hand side is a monotonically increasing function of $r^2$, starting from the origin and reaching $1$ at  $r=1$. Therefore, the left and the right hand sides must intersect at a single $r\in(0,1)$ (see Fig.\ref{r2 eq}a). In the special case when $R_1=R_2$, the right hand side is also a straight line and the interstection is at $1/2$. We can also clearly see that when $R_1>R_2$, the curve on the right hand side will sit below the straight line and therefore the solution will be larger than $1/2$; for $R_1<R_2$ it will be smaller. From this graphical solution we also see that the solution is a monotonic function of $R_1/R_2$, vanishing when $R_1/R_2\rightarrow 0$ and approaching $1$ when $R_2/R_1\rightarrow 0$. 
The numerical solution using the Newton's method is shown in Fig.\ref{r2 eq}b.

We can also eliminate $r$ in Eqs. (\ref{phi1}) and (\ref{eq:R2}) and obtain the generalized van der Pauw relation \cite{vanderPauw1958A}
\begin{eqnarray}
    e^{-\pi R_1/\rho_*}+e^{-\pi R_2/\rho_*}=1.
\end{eqnarray}
Therefore, the two resistances $R_1$ and $R_2$ allow us to determine $\rho_*$ and the hypergeometric parameter $r$.
    
Configuration 3 is a bit more subtle due to the appearance of logarithms of negative numbers. Pluggin in $z_S=0$, $z_D=1/r^2$, and $z=1$ in our expression for the potential \eqref{solution},
    \begin{equation}
        \Delta\Phi_3 =  \frac{I}{2\pi} \left[ \frac{1}{ \sqrt{\sigma_+\sigma_-}  + i \sigma_H} \ln{\frac{1-1/r^2}{1}} + \text{c.c.} \right].
    \end{equation}
    $r\in(0,1)$, so the argument in the logarithm is always negative. This is handled using the analytical continuation of the logarithm, 
        $\ln z = \ln\abs{z} + i\arg(z)$, but one must be careful of branch cuts at arguments of $\pi$ using limits. Since the Schwarz-Christoffel transformation maps points to the upper half of the complex plane, the contact, $A$, can be taken to be at $1+i\epsilon$, where $0<\epsilon\ll1$. The potential difference in configuration 3 is then written as \begin{equation}
        \Delta\Phi_3 =  \frac{I}{2\pi} \left[ \frac{1}{ \sqrt{\sigma_+\sigma_-}  + i \sigma_H} \ln{\frac{1+i\epsilon-1/r^2}{1+i\epsilon}} + \text{c.c.} \right].
    \end{equation}
    Simplifying the logarithm to linear order in $\epsilon$ and then taking the limit that $\epsilon\xrightarrow{}0$, $
        \ln(1-1/r^2+i\epsilon/r^2) = \ln(1/r^2-1) + i \pi.$
    The positive logarithm term  factors out, leading to a term that matches those in configurations 1 and 2,
but there is an additional term from the argument of the complex logarithm. After simplifying, we find
     \begin{eqnarray}
        R_3&&=\frac{\Delta\Phi_3}{I} =  \frac{\rho_*}{\pi} \ln(1/r^2-1) -\rho_H\\
       &&=R_2-R_1-\rho_H.
    \end{eqnarray}
    Where we defined $R_3$ keeping the original sign because, similar to the Hall resistance, its sign is physically relevant.
    Note that the Hall resistivity is 
    \begin{equation}
        \rho_H=-\frac{\sigma_H}{\sigma_H^2 + \sigma_+ \sigma_-}.
    \end{equation}
    Of four possible vertex configurations (see Fig.\ref{abstract}f), only three are linearly independent and provide new information. Plugging in $z_S=1$, $z_D=\infty$, $z_A = 0$, and $z_B=1/r^2$, the resistance for configuration 4 becomes (see Appendix \ref{config4}), \begin{equation}
         R_4 =  \frac{\rho_*}{\pi} \ln(1/r^2-1) +\rho_H,
    \end{equation} which differs from $R_3$ by the sign of the $\rho_H$ term.
    The fourth configuration acts as a check between the resistances measured from the other three vertex configurations, 
    \begin{equation}
         R_4 = -2 R_1 +  2 R_2 - R_3.
    \end{equation}

    With $R_1$, $ R_2$, and $R_3$ as defined above, we obtain three equations with three unknowns, $\sigma_H$, $ \sqrt{\sigma_+\sigma_-}$, and $r$. Therefore, only the geometric mean and the Hall conductivity can be determined with the three vertex configurations.  
    The first and second vertex configurations fix the hypergeometric parameter, $r$ \eqref{fixr}. With $r$ fixed, configurations 1 and 3 in Fig.\ref{abstract}f provide the Hall conductivity, \begin{equation} 
    \sigma_H =\frac{\ln(1-r^2)[R_3\ln(1-r^2) + R_1\ln(1/r^2-1)]}{(R_1\pi)^2 + [R_1\ln(1/r^2-1) + R_3\ln(1-r^2)]^2},
\end{equation} and the geometric mean,
\begin{equation} \label{geomean}
     \sqrt{\sigma_+ \sigma_-} = -\frac{R_1\pi\ln(1-r^2)}{(R_1\pi)^2 + [R_1\ln(1/r^2-1) + R_3\ln(1-r^2)]^2}.
\end{equation}
See Appendix \ref{derivehallandgeomean} for a derivation of our expressions for $\sigma_H$ and $\sqrt{\sigma_+\sigma_-}$.

Obtaining the orientation of the principal axes, i.e. the angle $\alpha$ and therefore the full conductivity tensor, requires an additional measurement not across vertices. To ensure the best spacing of contacts away from each other, while keeping them on the perimeter, we propose to measure at a midpoint. Although the analytical $z$-plane location of the physical midpoints after the Schwarz-Christoffel transformation is unknown, it can be determined using previous vertex resistance measurements.
The location of the midpoint, $z_5$, in the complex plane can be determined using configuration 5 (see Fig.\ref{abstract}g) in terms of the hypergeometric parameter $r$ and the resistance from configuration 1 \eqref{phi1}, 
\begin{align} 
    &R_5 = -\frac{R_1 }{\ln(1-r^2)}\ln\left(\frac{z_5-1/r^2}{z_5-1}\right)\nonumber
    \\
  \Rightarrow  &z_5 = \frac{1/r^2 - (1-r^2)^{-R_5/R_1}}{1- (1-r^2)^{-R_5/R_1}}.
\end{align} 
Equivalently, the midpoint can also be found in terms of $R_2$,
\begin{equation} \label{eq: midpoint}
    z_5 = 1-\frac{1/r^2-1}{(1/r^2)^{R_5/R_2} -1}.
\end{equation}
$R_5$ is a very similar measurement to $R_2$, but with the $A$ contact closer to the source. This makes $R_5$ larger than $R_2$. Since $0<r<1$, the denominator in Eq.\eqref{eq: midpoint} must be larger than the numerator. Therefore, $z_5$ cannot exceed $1$. As $r\rightarrow 1$, $z_5\rightarrow 1-\frac{R_2}{R_5}$. The denominator also grows with increasing $1/r^2$ faster than the numerator. Therefore, $1-\frac{R_2}{R_5}<z_5<1$.
 If $z_5$ is found to be outside of $(0,1)$, or equivalently $R_2>R_5$ then one of our starting assumptions must have been violated---for instance that conductivity tensor is spatially uniform.

In the w-plane, the midpoint of the lower edge is located at $\mathcal{K}_a(r)/2$. Therefore, using \eqref{sc}, this constraint becomes,
\begin{equation}
    \mathcal{K}_a(r)=\frac{z_5^{1-a}}{1-a} \sin (\pi a) F_1(1-a;1-a,a;2-a;z_5,r^2z_5).
\end{equation}
At $a=0$ or $a=1$ the function on the left hand side, $\mathcal{K}_a(r)$, is $\pi/2$ for any $r$. The right hand side is $0$ for $a=0$ and $\pi$ for $a=1$ for any $r\in (0,1)$ and any $z_5\in (0,1)$. Therefore, subtracting the left hand side from the right hand side yields a continuous function of $a$ with range $-\pi/2$ to $\pi/2$ on domain $a\in(0,1)$. By the intermediate value theorem, this must intersect $0$ at least once, guaranteeing a solution exists.
In practice we can easily solve the above equation numerically and find a unique solution for $a$ (see the provided \href{https://github.com/jag47/Magnetoconductance}{GitHub page}).

Knowing $a$, the principal axes angle $\alpha$ is found by combining \eqref{tana} and \eqref{ratio}. After much simplification (see Appendix \ref{ratioandalpha}), 
\begin{equation} \label{principle axes}
    \tan 2\alpha = \frac{2d_1d_2\mathcal{K}_a(r)\mathcal{K}_a(\sqrt{1-r^2})}{d_1^2\mathcal{K}_a(\sqrt{1-r^2})^2-d_2^2\mathcal{K}_a(r)^2} \cos a \pi.
\end{equation}
The left hand side is periodic with period $\pi/2$ and takes on the values from $-\infty$ to $\infty$ once on its period. Since the right hand side is a real number, which may be positive or negative, there are always two solutions for $\alpha\in[0,\pi)$. In order to decide which roots to take, we refer to equation \eqref{tana} which shows that with our choice of $\sigma_+\geq\sigma_->0$, if $0<a<1/2$, then $\alpha\in(0,\pi/2)$. Similarly, when $1/2<a<1$, then $\alpha\in(\pi/2, \pi)$. For the special case of $a=1/2$, $\alpha$ can be either $0$ or $\pi/2$. As we show next, the value of $r$ differentiates between these two special cases and pins down $\alpha$ uniquely.

When $a=1/2$, the constraint equation for $r$ \eqref{ratio} simplifies to \begin{align}\label{eq:ahalf_cases}
    \frac{K(r^2)}{K(1-r^2)} = \begin{cases}
        \frac{d_1}{d_2} \sqrt{\frac{\sigma_-}{\sigma_+}} \quad \text{if} \quad \alpha = 0
        \\ \frac{d_1}{d_2} \sqrt{\frac{\sigma_+}{\sigma_-}} \quad \text{if} \quad \alpha = \pi/2
    \end{cases},
\end{align} where $K(r^2)=\int_0^{\pi/2}d\theta/\sqrt{1-r^2\sin^2\theta}$ is the complete elliptic integral of the first kind.  As $r$ increases from $0$ to $1$, the left hand side of Eq.\eqref{eq:ahalf_cases} increases from $0$ to $\infty$, so the special case where $\sigma_+=\sigma_-$ separates the two scenarios for $\alpha$. Let this special value of $r$ be referred to as $r_{\text{isotropic}}$ and defined via the equation $d_2K\left(r^2_{\text{isotropic}}\right)=d_1K\left(1-r^2_{\text{isotropic}}\right)$. If $r$ determined by measured resistances using \eqref{fixr} is less than or equal to $r_{\text{isotropic}}$ then $\alpha=0$, otherwise $\alpha = \pi/2$. If $r=r_{\text{isotropic}}$ i.e. if the two principal value conductivities are the same, then $\alpha$ is physically meaningless since the entire conductivity tensor (including the Hall component) is isotropic. This then uniquely determines $\alpha$ given $a$ (and $r$).

Plugging in $\alpha$ found using \eqref{principle axes} into \eqref{tana}, solving the quadratic equation for  $\sqrt{\sigma_-/\sigma_+}$ and taking care to keep the physical positive root, we find
\begin{equation} \label{conratios}
    \sqrt{\frac{\sigma_-}{\sigma_+}}= \frac{-1+\sqrt{1+\tan^2(a\pi)\sin^2(2\alpha)}}{\tan(a\pi)\sin(2\alpha)}.
\end{equation}
The numerator is clearly positive semidefinite and based on the previous discussion of the choice of $a$ given $\alpha$, so is the denominator. Therefore the right hand side is guaranteed to be nonnegative, as it should be. Moreover, by the triangle inequality, the anisotropy ratio will always be less than or equal to one. This aligns with our initial assumption that $\sigma_+\geq \sigma_-$.

It is now trivial to get the diagonal conductivities: divide \eqref{geomean} and \eqref{conratios} to find $\sigma_+$; multiply \eqref{geomean} and \eqref{conratios} to find $\sigma_-$. The full conductivity tensor and principal axes angle are thus found by 4 potential measurements---3 vertex configurations, and 1 midpoint configuration. 
In practice, contacts can be placed at all four midpoints of the rectangle so that the experimental procedure above can be repeated for all four possible orientations of the sample, providing an error estimate.

To check our equations for finding materials parameters from resistance measurements, we simulated our four experimental configurations with COMSOL and extracted the conductivity tensor and anisotropy angle. Calculated materials parameters from simulated data using our equations agreed very well with the conductivity tensor and principal axes angle of the samples, verifying our provided formulas. 
     
\section{Discussion} \label{discussion}
    In this section we first discuss various checks of our solution by comparing to finite element methods and the limiting case where $\alpha=0$ which exists in the literature \cite{Simon1999Comment}. Then we generalize our solution using superposition to finite sized sources and drains along the perimeter of the sample.

\begin{figure*} 
        \centering        \includegraphics[width=1\linewidth]{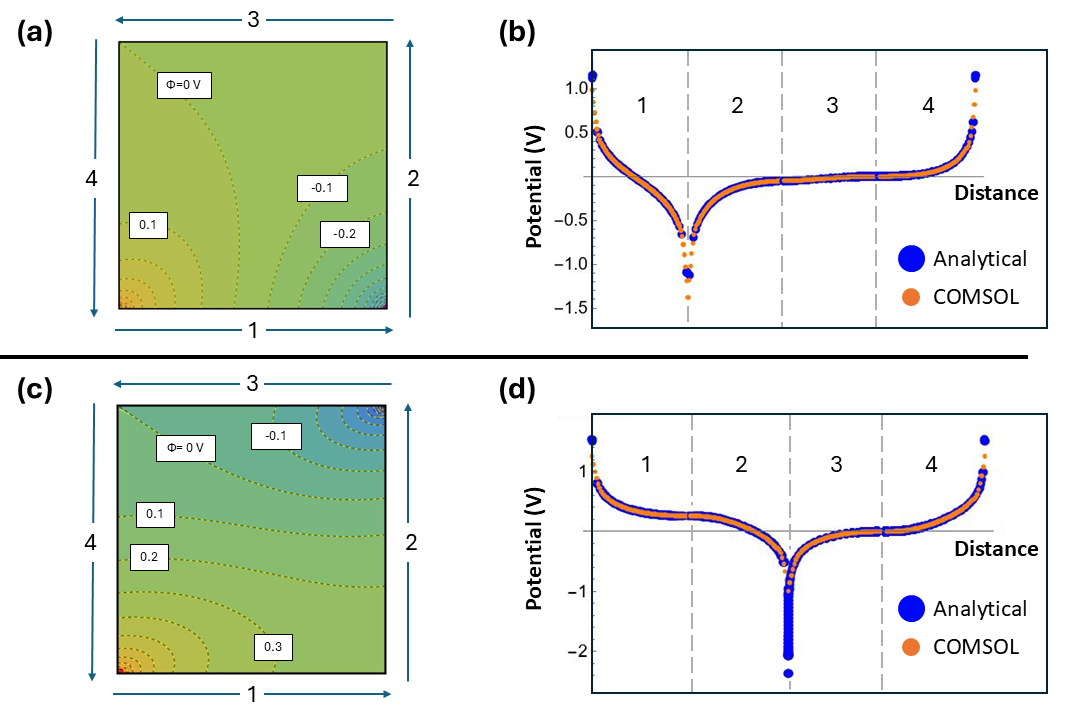}
        \caption{Comparison of our analytical solution with simulations from COMSOL. \textbf{(a)} Contour plot generated by COMSOL overlaid by dashed black contours generated by our solution. Here, $\sigma_+\approx 4.4 \text{ S}/\square, \sigma_-\approx 2.6 \text{ S}/\square, \sigma_H = 0.25 \text{ S}/\square$, and $\alpha \approx 28^\circ$. High potential is red (bottom left corner) and low potential is blue (bottom right corner). \textbf{(b)} 
        Counterclockwise edge cuts of potential from the contour plot (a). \textbf{(c)} $\sigma_+\approx 5.2 \text{ S}/\square, \sigma_-\approx 1.8 \text{ S}/\square, \sigma_H = -0.25 \text{ S}/\square$, and $\alpha \approx 13^\circ$. High potential is red (bottom left corner) and low potential is blue (top right corner). \textbf{(d)} 
        Potential on the edges of the contour plot (c) going counterclockwise. }
        \label{COMSOL}
\end{figure*}

    To verify the validity of our analytical solution for the potential, we compared our solution with one computed using COMSOL. See \ref{methods} for more details about the implementation of the numerical simulation. The data generated by COMSOL agreed very well with our analytical solution as shown in Fig.\ref{COMSOL}a and c. Along the edges of the sample, the analytical and numerical solutions are compared in Fig.\ref{COMSOL}b and d. Note that the conductivity tensors and anisotropy angles were changed between the two simulations shown, demonstrating the validity of the solution for different material parameters.

    To further verify our solution, we check that it limits to the known result when $\alpha=0$ (see Ref. \cite{Simon1999Comment}), in which case the Schwarz-Christoffel transformation is analytically invertible,
    \begin{equation}
        z = f(w) = \mathrm{csch}^2 \left[ -i*\mathrm{am}(-K(k)-iwr,k)  \right] +1.
    \end{equation}
    Here $\mathrm{am}(u,k)$ is the Jacobi amplitude function \cite{Weisstein2002Jacobi} with elliptic modulus $k = 1-1/r^2$, $K$ is a complete elliptic integral of the first kind, and $w$ is a coordinate in the rescaled parallelogram. This inverse transformation allows an explicit potential solution in terms of the original rectangle coordinates (see Eqs.\eqref{solution},\eqref{pointtransform} and \eqref{unirescale}),
    \begin{align} \label{phix}
        &\Phi(x,y) = \frac{I}{2\pi} \left[ \frac{1}{ \sqrt{\sigma_+\sigma_-}  + i \sigma_H} \ln{\frac{f(w)-f(w_D)}{f(w)-f(w_S)}} + \text{c.c.} \right]\nonumber
        \\
        &w_j = \frac{\mathcal{K}_a(r)}{x_1''} \left[R_{\theta}SR_{-\alpha} \begin{pmatrix}
        x \\  y \end{pmatrix}\right]_j, \quad w = w_1 + i w_2.
    \end{align}
   
\begin{figure}
        \centering
        \includegraphics[width=0.7\linewidth]{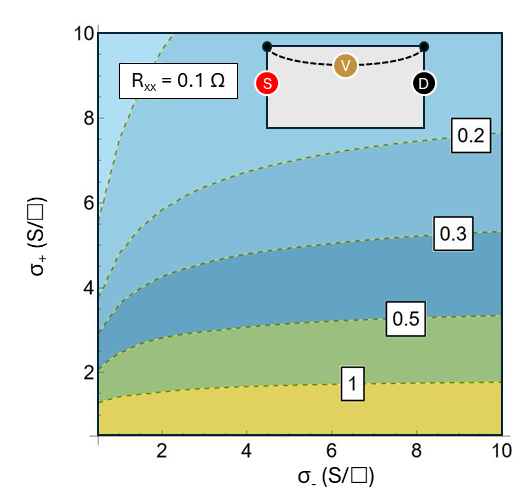}
        \caption{Contour plot of $R_{xx}$ calculated with our solution (shading and yellow contours) overlaid with $R_{xx}$ calculated using the sum in Ref.\eqref{simonsum} (black dashed contours) \cite{Simon1999Comment}. Inset shows the setup with source (S), drain (D), and voltage (V). Here $d_1/d_2=2.3/1.2$.}
        \label{simons}
    \end{figure}

    The result in reference \cite{Simon1999Comment} provides the longitudinal resistance along the x direction, assuming that the source/drain are at midpoints on the left/right sides of a rectangular sample (see the inset in Fig.\ref{simons}), 
    \begin{equation} \label{simonsum}
        \text{Ref \cite{Simon1999Comment}: }R_{xx} =\frac{4}{\pi} \frac{1}{\sqrt{\sigma_+\sigma_-}} \sum_{n,\text{odd}} \frac{1}{n} \mathrm{csch} \left( \sqrt{\frac{\sigma_+}{\sigma_-}} \frac{\pi n}{2} \frac{d_2}{d_1} \right).
    \end{equation}
    The resistance calculated by our solution (Eq.\eqref{phix}) gives 
    \begin{align}
         &\text{This paper: } R_{xx} =  \frac{1}{\pi}\frac{1}{\sqrt{\sigma_+ \sigma_-}} \ln \left[ \frac{F(r,k,C_1)}{F(r,k,C_2)} \right]\nonumber
        \\
        & F(r,k,C_j) =1 -r^2 + r^2 \mathrm{dc}^2\left(\frac{r}{2}K(r^2)C_j,k\right)\nonumber
        \\
        &C_1 = -\frac{d_2}{d_1}\sqrt{\frac{\sigma_+}{\sigma_-}},  \qquad C_2 = \frac{i*2d_1\sqrt{\sigma_-}-d_2\sqrt{\sigma_+}}{d_1\sqrt{\sigma_-}},
    \end{align}
    where $K$ is  a complete elliptic integral of the first kind, $r$ is a hypergeometric parameter (here an elliptic parameter) determined by \eqref{ratio}, and $\mathrm{dc}(u,k)$ is a Jacobi elliptic function \cite{WeissteinJacobi} with elliptic modulus $k = 1-1/r^2$. Plotting the two sums as functions of $\sigma_\pm$ reveal they are equivalent (see Fig.\ref{simons}b). Thus, our solution successfully reduces to the limiting case of vanishing anisotropy angle.

    \subsection{Multiple or Finite Size contacts}

    Since the continuity equation is linear, finite sources can be modeled with the principle of superposition. Using our solution \eqref{solution}, a potential from $N_S$ sources and $N_D$ drains at $z_{S/D,k}$ respectively each with current $\abs{I_k}$ is represented as

    \begin{align}
        \Phi(z) = \sum_{k=1}^{N_D} \frac{I_k} {2\pi}\left( \frac{\ln(z-z_{D,k})}{\sqrt{\sigma_+\sigma_- }+ i \sigma_H} + \mathrm{c.c}\right) &\\-\sum_{k=1}^{N_S} \frac{I_k} {2\pi}\left( \frac{\ln(z-z_{S,k})}{\sqrt{\sigma_+\sigma_- }+ i \sigma_H} + \mathrm{c.c}\right).
    \end{align}
    
    Taking $I_k$ to be infinitesimal, the Riemann sum can be converted to an integral. As an example, assume that the source and drain current density, $j = I/d$ is uniform in the lab frame. Furthermore, assume that the sample is square with side length $d$ and that the top side acts as a source and the bottom side as a drain. Hence, in the lab frame, $\dd{I} = \abs{j}\dd{x}$ along the bottom and top of the sample. After rotating and rescaling \eqref{pointtransform}\eqref{unirescale},  $\dd{x} = \frac{d}{\mathcal{K}_a(r)}\dd{w}$. Finally, after conformal mapping \eqref{sc}, \begin{equation}
        \dd{w} = \abs{f'(z)}\dd{z} = \frac{\sin \pi a}{2} \abs{  \frac{1}{z^a (1-z)^{1-a}(1-r^2z)^a}}\dd{z}.
    \end{equation}

\begin{figure}
        \centering
    \includegraphics[width=0.7\linewidth]{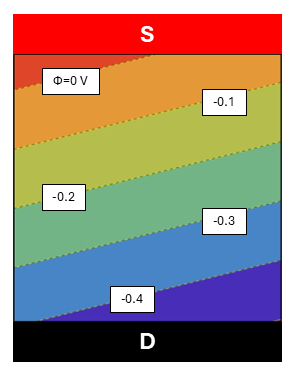}
        \caption{
        Contour plot of the potential for a square sample with a continuous source along the top and a drain along the bottom with uniform current density, $\abs{j}\approx1.08 \text{ A/m}$. For this sample, $\sigma_+\approx 4.4 \text{ S}/\square$, $ \sigma_-\approx 2.6 \text{ S}/\square$, $ \sigma_H = 0.25 \text{ S}/\square $ and $\alpha \approx 28^\circ$. Shading and yellow contours are from COMSOL simulations, while the black dashed lines are generated from superposition of our analytical solution \eqref{eq: finite potentail}.}
        \label{finitecontacts}
    \end{figure}

    Combining how the current density transforms from the square to the complex plane, the potential is then written as, 
    \begin{align} \label{eq: finite potentail}
        \Phi(z) &= \frac{I}{2\pi}\frac{1}{\mathcal{K}_a(r)} \Biggl[ \int_0^1 \dd{z'}\abs{f'(z')}\left( \frac{\ln(z-z')}{\sqrt{\sigma_+\sigma_- }+ i \sigma_H} + \mathrm{c.c}\right)\nonumber
        \\
        &-\int_{1/r^2}^\infty\dd{z'}\abs{f'(z')}\left( \frac{\ln(z-z')}{\sqrt{\sigma_+\sigma_- }+ i \sigma_H} + \mathrm{c.c}\right)\Biggr].
    \end{align}
    Numerical integration of the above equation agrees very well with numerical simulations in COMSOL (see Figure \ref{finitecontacts}).

\section{Summary}
We developed an analytical solution for the electrical potential of a uniform 2D rectangular sample with contacts on the perimeter using conformal mapping techniques. Using our solution, the full conductivity tensor (including Hall components) and principal axes angle can be extracted simply from four measurements. Numerical simulations in COMSOL agreed very well with our analytical solution.  

\section{Acknowledgments}
The authors would like to express their sincere gratitude to Prof. Michael Zudov for numerous discussions and for motivating this project. OV would also like to thank Prof. Matti Vuorinen for helpful correspondence. This work was partially supported by the Research Experiences for Undergraduates (REU) Program of the National Science Foundation under Award Number PHY-2348668; OV is also supported by Simons Foundation under Grant No. SFI-MPS348 NFS00006741-09.

\section{Methods} \label{methods}
    COMSOL simulations were performed under the Poisson equation interface with a stationary study. One point source and one point drain were placed at the corners of the sample as specified carrying current, $\abs{I} = 1.3 \text{ A}$. A Dirichlet boundary condition was placed at the top left corner to ground the sample as in our solution. For finite sized contacts, a flux/source was added to the top and bottom of the sample with current density, $\abs{j} = 1.3 \text {A}/1.2  \text { m}\approx 1.08 \text{ A/m} $. $\sigma_+\approx 4.4 \text{ S}/\square, \sigma_-\approx 2.6 \text{ S}/\square, \sigma_H = 0.25 \text{ S}/\square$ corresponds to a lab frame conductivity tensor of $\sigma = \begin{pmatrix}
        4 && 1\\ 0.5 && 3
    \end{pmatrix} \text{ S}/\square$ which was entered into COMSOL for Fig.\ref{COMSOL}a and b and \ref{finitecontacts}. $\sigma_+\approx 5.2 \text{ S}/\square, \sigma_-\approx 1.8 \text{ S}/\square, \sigma_H = -0.25 \text{ S}/\square$ corresponds to $\sigma = \begin{pmatrix}
        5 && 0.5\\ 1 && 2
    \end{pmatrix} \text{ S}/\square$ used in Fig.\ref{COMSOL}c and d.

\bibliography{refs}

\appendix

\section{Fourier Transformation Solution} \label{fullsolution}
    The solution to the differential equation \eqref{diffeq} with boundary conditions \eqref{complex bounds} is found using Fourier transformations. Assume that $\Phi(z)$ is of the form
    \begin{equation} \label{fourier}
        \Phi(z=u+iv)  = \int e^{i k u} \Phi_k (v).
    \end{equation}

    Plugging \eqref{fourier} into \eqref{diffeq} and using the orthogonality of Fourier modes, $\Phi_k$ yields an ordinary differential equation in $v$, 
    \begin{equation}
        \sqrt{\sigma_+\sigma_-}\left(k^2 -\pdv[2]{}{v}\right)\Phi_k = \frac{I}{2\pi} \delta(v-v_{\epsilon}) (e^{-i k u_{S}} - e^{-i k u_{D}})
    \end{equation}
    Integrating this equation from $v_\epsilon-\eta$ to $v_\epsilon+\eta$ and taking the limit that $\eta \xrightarrow{}0$,
    \begin{equation} \label{constraint 2}
    \sqrt{\sigma_+\sigma_-}\left( \pdv{\Phi_k^<}{v} |_{v_{\epsilon}} - \pdv{\Phi_k^>}{v}|_{v_{\epsilon}}\right) = \frac{I}{2\pi} (e^{-i k u_{S}} - e^{-i k u_{D}}),
\end{equation}
    where $\Phi_k^>$ is defined for $v>v_{\epsilon}$ and $\Phi_k^< $ is defined for $0 \leq v\leq v_{\epsilon}$. Requiring no divergences for large $v$,
\begin{equation}
\begin{split} \label{phi}
    &\Phi_k^> = C_k e^{-|k|v}
    \\
    &\Phi_k^< = A_k e^{kv} + B_k e^{-kv}.
\end{split}
\end{equation}

The undetermined constants are found by demanding continuity at $v_\epsilon$ (1), the continuity equation \eqref{constraint 2} to hold (2), and the boundary conditions \eqref{complex bounds} to be met (3):

\begin{equation}
    \begin{split}
        &1. \quad A_k e^{kv_\epsilon} + B_k e^{-kv_\epsilon} = C_k e^{-|k|v_\epsilon}
        \\
        &2. \quad \sqrt{\sigma_+\sigma_-} k \left( A_k e^{kv_\epsilon} - B_k e^{-kv_\epsilon} + \frac{k}{|k|} C_k e^{-|k|v_\epsilon} \right) \\ &\qquad = \frac{I}{2\pi} (e^{-i k u_S} - e^{-i k u_D})
        \\
        &3. \quad \sqrt{\sigma_+\sigma_-}(B_k-A_k)+i \sigma_H(A_k+B_k)=0.
    \end{split}
\end{equation}
Solving the above for $C_k$ and taking the limit that $v_{\epsilon}\xrightarrow{}0$,
\begin{equation} \label{ck}
    C_k = \frac{I}{2\pi} \frac{e^{-i k u_S} - e^{-i k u_D}}{i k \sigma_H + \abs{k} \sqrt{\sigma_+\sigma_-}}.
\end{equation}
With $C_k$ determined, $\Phi$ is written as an integral of Fourier modes, 
\begin{equation}
    \Phi(z) = \frac{I}{2\pi}\int_{-\infty}^{\infty} \dd{k} e^{i k u-\abs{k}v}  \frac{e^{-i k u_S} - e^{-i k u_D}}{i k \sigma_H + \abs{k} \sqrt{\sigma_+\sigma_-}} 
\end{equation}
The absolute value is handled by breaking the integral into two, with one being the complex conjugate of the first, and changing the bounds from $0$ to $\infty$. Evaluating this integral yields our solution for the potential in the complex plane:
\begin{equation} 
    \Phi(z) = \frac{I}{2\pi} \left[ \frac{1}{ \sqrt{\sigma_+\sigma_-}  + i \sigma_H} \ln{\frac{z-z_D}{z-z_S}} + \text{c.c.} \right].
\end{equation}

\section{Midpoint Cross-check} \label{midpointcross}
    Many additional cross-checks can also be performed using midpoints of the sample after determining the hypergeometric parameters. Here is an example using variations of configuration 5 (see Fig.\ref{abstract}g). 
    
    The analytical locations of the midpoints int the complex plane are unknown, but they can be determined using $R_1$ and $r$, a hypergeometric parameter. In configuration 5, the location of $A$ in the complex plane, $z_{\text{bottom}}$, is given by 
    \begin{equation} \label{eq: bottommid}
        z_5 = z_{\text{bottom}} = \frac{1/r^2 - (1-r^2)^{-R_{\text{bottom}}/R_1}}{1- (1-r^2)^{-R_{\text{bottom}}/R_1}},
    \end{equation}
    where $R_{\text{bottom}}=R_5$ is the measured resistance. 
    
    Consider a similar configuration with $B$ at the top left corner, $A$ at the center of the top edge, and the source and drain at the same locations as in configuration 5. The location of $A$, the top midpoint, in the complex plane is given by
    \begin{equation}
        z_{\text{top}} = \frac{1/r^2 - (1-r^2)^{R_{\text{top}}/R_1}}{1- (1-r^2)^{R_{\text{top}}/R_1}}.
    \end{equation}
    Note the change in the sign of the exponent, $R_{top}/R_1$, compared to \eqref{eq: bottommid}. We define $R_{top}$ to be positive which requires a negative sign since the top midpoint is closer to the drain than the top left corner.

    Using the found location of these two midpoints in the complex plane, the resistance across the two midpoints can be predicted as a cross-check. Keeping the source and drain as pictured in configuration 5, the resistance across the two midpoints above is expected to be 
    \begin{equation}
    R_{\text{midpoints}}=-\frac{R_1}{\ln(1-r^2)} \ln\frac{(z_{\text{bottom}}-1/r^2)(z_{\text{top}}-1)}{(z_{\text{bottom}}-1)(z_{\text{top}}-1/r^2)}
    \end{equation} when measured from the bottom midpoint ($A$) to the top midpoint ($B$).

\section{Resistance for Configuration 4}\label{config4}
    Similar to vertex configuration 3, the potential in configuration 4 contains complex logarithms, but it also has infinities that must be handled carefully. Plug in $z_S=1$, $z_D=\infty$, into our expression for the potential and compute the difference between the potential at $A=0$ and $B=1/r^2$,
    \begin{align}
        \Delta\Phi_4 = \frac{I}{2\pi} \Biggl[ \frac{1}{\sqrt{\sigma_+\sigma_-} + i\sigma_H} \ln \frac{0-\infty}{0-1}  + \mathrm{c.c.}\nonumber \\  -\left(\frac{1}{\sqrt{\sigma_+\sigma_-} + i\sigma_H} \ln \frac{1/r^2-\infty}{1/r^2-1}  + \mathrm{c.c.} \right)\Biggr ].
    \end{align}
    The logarithms individually diverge since they are evaluated at $\infty$, but their difference is finite,
    \begin{align}
        \Delta\Phi_4 &= \frac{I}{2\pi} \Biggl[ \frac{1}{\sqrt{\sigma_+\sigma_-} + i\sigma_H} \ln \frac{(0-\infty)(1/r^2-1)}{(0-1)(1/r^2-\infty)}  + \mathrm{c.c.}]\nonumber
        \\&
        = \frac{I}{2\pi} \Biggl[ \frac{1}{\sqrt{\sigma_+\sigma_-} + i\sigma_H} \ln (1-1/r^2)  + \mathrm{c.c.}\Biggr].
    \end{align}
    Since $r\in(0,1)$, the logarithm acts on a negative number.  It must be computed with its analytical continuation, $\ln z  = \ln \abs{z} + i\arg z$, keeping careful note of branch cuts. Consider the limit where $B$ is placed just slightly off the perimeter of the sample so that $B = 1/r^2 + i\epsilon$ ( $0<\epsilon\ll 1$). Taking the limit that $\epsilon\xrightarrow[]{}0$, the logarithm becomes $\ln(1-(1/r^2 + i \epsilon)) = \ln(1/r^2-1) -i\pi $.

    The expression for $\Delta\Phi_4$ has the same form as $\Delta\Phi_3$ except the sign of the argument in the complex logarithm changes,
    \begin{equation}
        \Delta\Phi_4 =  \frac{I}{2\pi} \left[ \frac{1}{ \sqrt{\sigma_+\sigma_-}  + i \sigma_H}\left( \ln(1/r^2-1) -i\pi\right) + \text{c.c.} \right].
    \end{equation}
    Thus, the resistance for configuration 4 is,
    \begin{equation}
        R_4=\frac{\Delta\Phi_4}{I} =  \frac{\rho_*}{\pi} \ln(1/r^2-1) +\rho_H.
    \end{equation}

\section{Deriving Expressions for the Geometric Mean and Hall Conductivity} \label{derivehallandgeomean}

    The geometric mean, $\sqrt{\sigma_+\sigma_-}$, and the Hall conductivity, $\sigma_H$, follow by straightforward algebra from our expressions for $R_1$ and $R_3$,
    \begin{align}
        &R_1 = - \frac{1}{\pi} \frac{\sqrt{\sigma_+ \sigma_{-}}}{\sigma_{H}^{2} + \sigma_+ \sigma_{-}} \ln(1 - r^{2})  
        \\&  R_{3} = \frac{1}{\pi} \frac{\sqrt{\sigma_+ \sigma_{-}}}{\sigma_{H}^{2} + \sigma_+ \sigma_{-}} \ln(1/r^{2} - 1) + \frac{\sigma_{H}}{\sigma_{H}^{2} + \sigma_+ \sigma_{-}}.
    \end{align}
    
    Rearrange the equations for $R_1$ and $R_3$ so that they both equal $\sigma_H^2 + \sigma_+\sigma_-$,
    \begin{align}
&\sigma_{H}^{2} + \sigma_+ \sigma_{-} = - \frac{1}{\pi R_{1}} \sqrt{\sigma_+ \sigma_{-}} \ln(1 - r^{2}) \label{plugin} \\& \sigma_{H}^{2} + \sigma_+ \sigma_{-} = \frac{\frac{1}{\pi} \sqrt{\sigma_+ \sigma_{-}} \ln(1/r^{2} - 1) + \sigma_{H}}{R_{3}},
\end{align}
    and then solve for $\sigma_H$ in terms of $\sqrt{\sigma_+ \sigma_{-}}$,
 \begin{equation}\label{sigmah}
\sigma_{H}=- \frac{\sqrt{\sigma_+ \sigma_{-}}}{\pi} \left(\frac{R_{3}}{ R_{1}} \ln(1 - r^{2}) +  \ln(1/r^{2} - 1)\right).
\end{equation}
Plugging the above expression for $\sigma_H$ into \eqref{plugin} and solving for $\sqrt{\sigma_+\sigma_-}$ yields our given expression for $\sqrt{\sigma_+\sigma_-}$,
    \begin{equation} \label{eq: geomean}
        \sqrt{\sigma_+ \sigma_-} = -\frac{R_1\pi\ln(1-r^2)}{(R_1\pi)^2 + [R_1\ln(1/r^2-1) + R_3\ln(1-r^2)]^2}.
    \end{equation}
    Finally, plugging \eqref{eq: geomean}  into \eqref{sigmah} yields our expression for $\sigma_H$,
    \begin{equation}
         \sigma_H =\frac{\ln(1-r^2)[R_3\ln(1-r^2) + R_1\ln(1/r^2-1)]}{(R_1\pi)^2 + [R_1\ln(1/r^2-1) + R_3\ln(1-r^2)]^2}
    \end{equation}

    \section{Deriving Expressions for the Ratio of Conductivity and Anisotropy Angle} \label{ratioandalpha}

    Our expressions for $\sqrt{\frac{\sigma_-}{\sigma_+}}$ and $\alpha$  also follow by algebra, but one must be careful of the sign of roots taken. First, rearrange the constraint equation for $a$ to get a quadratic equation in $\sqrt{\frac{\sigma_-}{\sigma_+}}$,
    \begin{align}
&\tan a\pi = \frac{\sqrt{\sigma_+ \sigma_-}}{\sigma_+ - \sigma_-} \frac{1}{\sin\alpha \cos\alpha}
\\&\tan a\pi \left(1 - \frac{\sigma_-}{\sigma_+}\right) - \sqrt{\frac{\sigma_-}{\sigma_+}} \frac{1}{\sin\alpha \cos\alpha} = 0.
\end{align}
The quadratic formula yields,
\begin{align}
\sqrt{\frac{\sigma_-}{\sigma_+}} = \frac{\frac{1}{\sin\alpha \cos\alpha} \pm \sqrt{\frac{1}{\sin^2\alpha \cos^2\alpha} + 4(\tan^2 a\pi)}}{- 2 \tan a\pi}.
\end{align} One obtains our expression for the anisotropy ratio by using the double angle formula for $\sin(2\alpha)$ and only keeping the physical positive root to ensure that $\sqrt{\frac{\sigma_-}{\sigma_+}}$ is greater than $0$,
\begin{equation} \label{ratioappend}
    \sqrt{\frac{\sigma_-}{\sigma_+}} = \frac{-1 + \sqrt{1 + \sin^2(2\alpha) \tan^2 a\pi}}{\sin(2\alpha) \tan a\pi}.
\end{equation} 

This gives $\sqrt{\frac{\sigma_-}{\sigma_+}}$ as a function of $\alpha$ since $a$ has already been fixed by a resistance measurement in configuration $5$ (see Fig.\ref{abstract}g). To find $\alpha$, use the original constraint equation for $r$, which is already known from vertex configurations 1 and 2 (see Fig.\ref{abstract}f),
\begin{equation}
\frac{\mathcal{K}_a(r)}{\mathcal{K}_a(\sqrt{1-r^2})} = \frac{d_1}{d_2} \sqrt{\frac{\sigma_- \cos^2 \alpha + \sigma_+ \sin^2 \alpha}{\sigma_+ \cos^2 \alpha + \sigma_- \sin^2 \alpha}}.
\end{equation}
Use this equation to solve for  $\frac{\sigma_-}{\sigma_+}$ as a function of $\alpha$,
\begin{align}
  &\frac{\frac{\sigma_-}{\sigma_+} + \tan^2 \alpha}{1 + \frac{\sigma_-}{\sigma_+} \tan^2 \alpha} =A  \quad \xrightarrow{} \quad  \frac{\sigma_-}{\sigma_+} = \frac{\tan^2 \alpha - A}{A \tan^2 \alpha - 1} \label{ratiosquared}
  \\&
  \text{ where }  A= \left[ \frac{d_2 \mathcal{K}_a(r)}{d_1 \mathcal{K}_a(\sqrt{1-r^2})} \right]^2 \label{Adef}.
\end{align}
To solve for $\alpha$ in terms of $a$ and $r$ only, substitute $\sigma_-/\sigma_+$ from the expression above into Eq.\eqref{ratioappend},
\begin{equation}
\frac{\tan^2 \alpha - A}{A \tan^2 \alpha - 1} = \left[ \frac{-1 + \sqrt{1 + \sin^2(2\alpha) \tan^2(a\pi)}}{\sin(2\alpha) \tan(a\pi)} \right]^2.
\end{equation}

The previous equation constrains $\alpha$ because $a$ and $r$ are already known. It will eventually become our given expression in terms of $\tan2\alpha$. To see this, first isolate the remaining square root term and square the entire expression,
\begin{align}
\Biggl[ -\frac{1}{2} \Biggl( \frac{(\tan^2 \alpha - A)\sin^2(2\alpha) \tan^2(a\pi)}{A \tan^2 \alpha - 1} - 2\nonumber \\- \sin^2(2\alpha) \tan^2(a\pi) \Biggr) \Biggr]^2 = 1 + \sin^2(2\alpha) \tan^2(a\pi)
\end{align}
After much simplification, the right hand side is best handled by completing the square,
\begin{align}
\sin^2(2\alpha) \tan^2(a\pi) (1-A)^2 (\tan^2\alpha + 1)^2 \nonumber \\ = 4(A(\tan^2\alpha + 1)^2  - \tan^2\alpha (1+A)^2). 
\end{align}

This yields an expression for $\sin2\alpha$ in terms of $a$ by using trig identities,
\begin{equation}\label{sin2a}
\sin^2(2\alpha)  = \frac{4A}{\tan^2(a\pi) (1-A)^2 + (1+A)^2}.
\end{equation}

This is nearly our given expression for $\tan2\alpha$. To turn this into an expression for $\tan2\alpha$, use the identity, \begin{equation*}
\sin^2(2\alpha) = \frac{\tan^2(2\alpha)}{1 + \tan^2(2\alpha)},
\end{equation*}
and substitute it into Eq.\eqref{sin2a},
\begin{equation}
\tan^2(2\alpha) = \frac{4A}{(\tan^2(a\pi) + 1)(1-A)^2}.
\end{equation}
We define $\alpha$ to be the positive angle between $0$ and $\pi$ measured counterclockwise from the x-axis when $\sigma_+\geq\sigma_-$, so only the positive square root solution is relevant,
\begin{equation}
\tan(2\alpha) =  \frac{2\sqrt{A}}{(1-A)} \cos(a\pi).
\end{equation}

Plugging $A$ \eqref{Adef} back into the above equation yields our given expression for $\tan2\alpha$,
\begin{equation}
    \tan 2\alpha = \frac{2d_1d_2\mathcal{K}_a(r)\mathcal{K}_a(\sqrt{1-r^2})}{d_1^2\mathcal{K}_a(\sqrt{1-r^2})^2-d_2^2\mathcal{K}_a(r)^2} \cos a \pi.
\end{equation}
With $\alpha$ thus determined, \eqref{ratioappend} determines $\sqrt{\frac{\sigma_-}{\sigma_+}}$ in terms of known variables, $a$ and $\alpha$.

\end{document}